\documentclass[twocolumn]{aastex631} %linenumbers
\usepackage{CJK}
\usepackage{amsmath} % convert program listings to a form includable in a LaTeX document

\usepackage{xcolor} % for bg color
\definecolor{bg-green}{rgb}{0.8588,0.9333,0.8666}
\definecolor{Q-color}{rgb}{0.5,0.1,0.5}
\definecolor{HHcolor}{rgb}{0.93,0.57,0.13}

\newcommand{\kms}[0]{{\,{\rm km\,s}^{-1}}}
\newcommand{\ms}[0]{{\,M_\odot}}
\newcommand{\ergs}[0]{{\,{\rm erg\,s}^{-1}}}

\shorttitle{Supersonic Turbulence in Galaxy Clusters}
\shortauthors{Hu et al.}

\begin{document} 
\begin{CJK*}{UTF8}{gbsn} 

\title{Signature of Supersonic Turbulence in Galaxy Clusters Revealed by AGN-driven H$\alpha$ Filaments}

\correspondingauthor{Yu Qiu}
\email{yuqiu@pku.edu.cn}

\author[0000-0003-3143-3995]{Haojie Hu (胡豪杰)}
\affiliation{Kavli Institute for Astronomy and Astrophysics, Peking University, 5 Yiheyuan Road, Haidian District, Beijing 100871, PRC}
\affiliation{Department of Astronomy, School of Physics, Peking University, 5 Yiheyuan Road, Haidian District, Beijing 100871, PRC}

\author[0000-0002-6164-8463]{Yu Qiu (邱宇)}
\affiliation{Kavli Institute for Astronomy and Astrophysics, Peking University, 5 Yiheyuan Road, Haidian District, Beijing 100871, PRC}

\author[0000-0002-7326-5793]{Marie-Lou Gendron-Marsolais}
\affiliation{Instituto de Astrof\'isica de Andaluc\'ia (IAA-CSIC), Glorieta de la Astronom\'ia, 18008 Granada, Spain}
\affiliation{European Southern Observatory, Alonso de C\'ordova 3107, Vitacura, Casilla 19001, Santiago, Chile}

\author[0000-0002-7835-7814]{Tamara Bogdanovi\'c}
\affiliation{Center for Relativistic Astrophysics, School of Physics, Georgia Institute of Technology, 837 State Street, Atlanta, GA 30332, USA}

\author[0000-0001-7271-7340]{Julie Hlavacek-Larrondo}
\affiliation{D\'epartement de Physique, Universit\'e de Montr\'eal, C.P. 6128, Succ. Centre-Ville, Montr\'eal, QC H3C 3J7, Canada}

\author[0000-0001-6947-5846]{Luis C. Ho}
\affiliation{Kavli Institute for Astronomy and Astrophysics, Peking University, 5 Yiheyuan Road, Haidian District, Beijing 100871, PRC}
\affiliation{Department of Astronomy, School of Physics, Peking University, 5 Yiheyuan Road, Haidian District, Beijing 100871, PRC}

\author[0000-0001-9840-4959]{Kohei Inayoshi}
\affiliation{Kavli Institute for Astronomy and Astrophysics, Peking University, 5 Yiheyuan Road, Haidian District, Beijing 100871, PRC}

\author[0000-0002-2622-2627]{Brian R. McNamara}
\affiliation{Department of Physics and Astronomy, University of Waterloo, 200 University Avenue West, Waterloo, ON, N2L 3G1, Canada}
\affiliation{Waterloo Center for Astrophysics, University of Waterloo, 200 University Avenue West, Waterloo, ON, N2L 3G1, Canada}
\affiliation{Perimeter Institute for Theoretical Physics, 31 Caroline St N, Waterloo, ON, N2L 2Y5, Canada}

%%%%%%%%%%%%%%%%%%%%%%%%%%%%%%%%%%%%%%%%%%%%%%%%%%%%%%%%%%%%%%%%%%%%%%%%%%%%%%%%%%%%%%%%%%%%%%%%%%%%%%%%%%%

\begin{abstract}
The hot intracluster medium (ICM) is thought to be quiescent with low observed velocity dispersions. Surface brightness fluctuations of the ICM also suggest that its turbulence is subsonic with a Kolmogorov scaling relation, indicating that the viscosity is suppressed and the kinetic energy cascades to small scales unscathed. However, recent observations of the cold gas filaments in galaxy clusters find that the scaling relations are steeper than that of the hot plasma, signaling kinetic energy losses and the presence of supersonic flows. In this work we use high-resolution simulations to explore the turbulent velocity structure of the cold filaments at the cores of galaxy clusters. Our results indicate that supersonic turbulent structures can be ``frozen'' in the cold gas that cools and fragments out of a fast, $\sim10^7$\,K outflow driven by the central active galactic nucleus (AGN), when the radiative cooling time is shorter than the dynamical sound-crossing time. After the cold gas formation, however, the slope of the velocity structure function (VSF) flattens significantly over short, $\sim10$\,Myr timescales. The lack of flattened VSF in observations of H$\alpha$ filaments indicates that the H$\alpha$-emitting phase is short-lived for the cold gas in galaxy clusters. On the other hand, the ubiquity of supersonic turbulence revealed by observed filaments strongly suggests that supersonic outflows are an integral part of AGN-ICM interaction, and that AGN activity plays a crucial role at driving turbulence in galaxy clusters.
\end{abstract}


%%%%%%%%%%%%%%%%%%%%%%%%%%%%%%%%%%%%%%%%%%%%%%%%%%%%%%%%%%%%%%%%%%%%%%%%%%%%%%%%%%%%%%%%%%%%%%%%%%%%%%%%%%%
\section{Introduction} \label{sec:intro}

In a turbulent fluid, kinetic energy ($e_{\rm kin}$) is expected to cascade from the driving scale to the viscous dissipation scale at a constant rate, i.e., $\mathrm{d} e_{\mathrm{kin}}/{\mathrm{d} t} \propto \rho v^{2}/{t} \propto \rho v^{3}/{l}\, =\,\text {const}$, where $\rho$, $v$, and $l$ are respectively the density, velocity, and length scale of the fluid elements, and $t$ is the energy transfer timescale. Under this assumption, a scaling relation can be drawn between $\rho$, $v$, and $l$, i.e., $\rho^{1/3}v\propto l^{1/3} $. This simple dimensional analysis offers profound insights into the turbulent properties of an incompressible fluid, where $v\propto l^{1/3} $ \citep{Kolmogorov1941}. Such a relation is found in X-ray observations of nearby galaxy clusters, where the one-component velocity amplitude scales with wavenumber ($k\equiv l^{-1}$) to the $-1/3$ power \citep{Zhuravleva2014, Zhuravleva2019}. Additionally, X-ray line widths have been used to probe the level of turbulence in galaxy clusters, e.g., \citet{Sanders2011} examined a sample of 62 systems, half of which have line width upper limits below $700\kms$. Recent X-ray calorimeter observations also found a low line-of-sight velocity dispersion (Gaussian $\sigma$) of $164\pm10\kms$ in the core of the Perseus cluster \citep{Hitomi2016}. Both of these observational results indicate that the level of turbulence in the ICM is mild and subsonic in a relaxed galaxy cluster.

The subsonic structure of the ICM does not preclude the presence of supersonic flows in galaxy clusters, particularly the outflows driven by the central AGN, which are believed to heat the surrounding medium and quench star formation \citep{Fabian2012, McNamara2012}. Recent spectral analysis of the spiral galaxy M81 provides direct evidence for hot, AGN-driven winds with line-of-sight velocities of $\pm2800\kms$, based on the Fe\,$\textsc{xxvi}$ Ly$\alpha$ line shifts \citep{Shi2021}.\footnote{Note that the wind in this low-luminosity AGN is likely from a hot accretion flow, and not from a radiatively efficient accretion disk. Both a wind and a mass-loaded relativistic jet may drive the outflows explored in this work \citep[for a discussion of these outflow-driving mechanisms, see][]{Qiu2021}.} In a rich cluster environment, however, direct detection of such outflows is more difficult due to the X-ray emission from the massive ICM. On the other hand, recent simulations of galaxy clusters and AGN feedback indicate that cold gas may fragment out of a supersonic, radiatively cooling outflow \citep{Qiu2020, Qiu2021a}, giving rise to extended filamentary H$\alpha$ nebulae, such as observed in the Perseus cluster \citep{Conselice2001, GM2018}. Therefore, the velocity structure of the $10^4$\,K gas offers a unique opportunity for probing the properties of the original hot outflow in galaxy clusters. 

%%%%%%%%%%%%%%%%%%%%%%%%%%%%%%%%%%%%%%%%%%%%%%%%%%
% FIGURE 1 simulation
%%%%%%%%%%%%%%%%%%%%%%%%%%%%%%%%%%%%%%%%%%%%%%%%%%
\begin{figure*}[t!]
\centering
\includegraphics[width=1\linewidth]{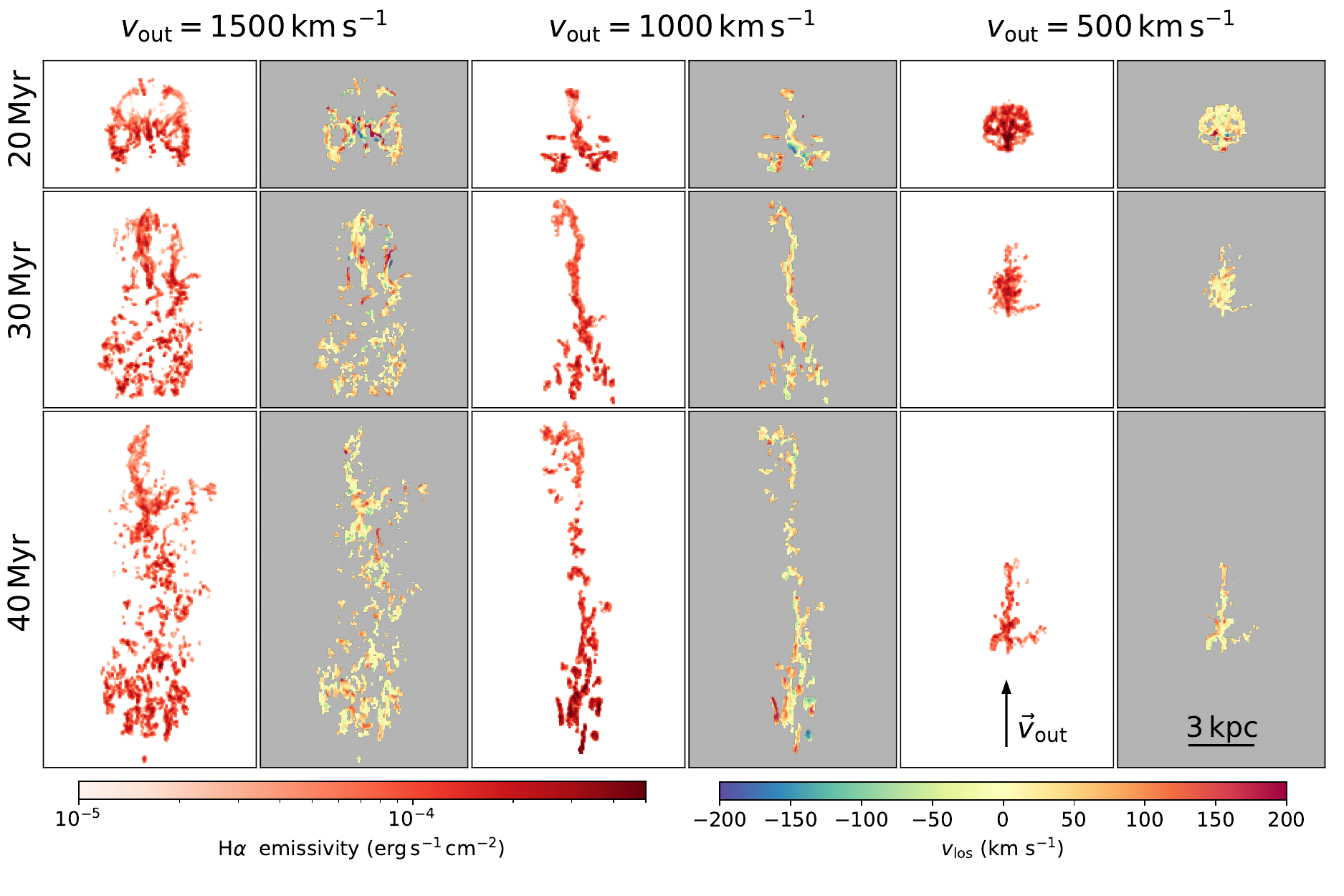}
\caption{The spatial distribution of the projected H$\alpha$ emissivity (columns 1, 3, 5) and the emissivity-weighted line-of-sight velocity (columns 2, 4, 6) for the three simulation runs with different initial outflow velocity, $v_{\rm out}$, projected along $\theta=90^\circ$ (viewing angle $\theta$ between sightline and initial outflow direction). The initial outflow direction and the length scale are shown in the two bottom right panels. The top, middle, and bottom rows show three representative evolution epochs at $t=20$, $30$, and $40\,$Myr, respectively. A horseshoe-shaped outer shell consisting of H$\alpha$-emitting gas is produced for the outflow with $v_{\rm out}=1500\kms$ at $t=20$\,Myr \citep[albeit smaller in size compared to the Perseus horseshoe filament;][]{Conselice2001}.}
\label{fig:projections}
\end{figure*}

%%%%%%%%%%%%%%%%%%%%%%%%%%%%%%%%%%%%%%%%%%%%%%%%%%
% FIGURE 2 simulation
%%%%%%%%%%%%%%%%%%%%%%%%%%%%%%%%%%%%%%%%%%%%%%%%%%
\begin{figure*}[t!]
\centering
\includegraphics[width=\linewidth]{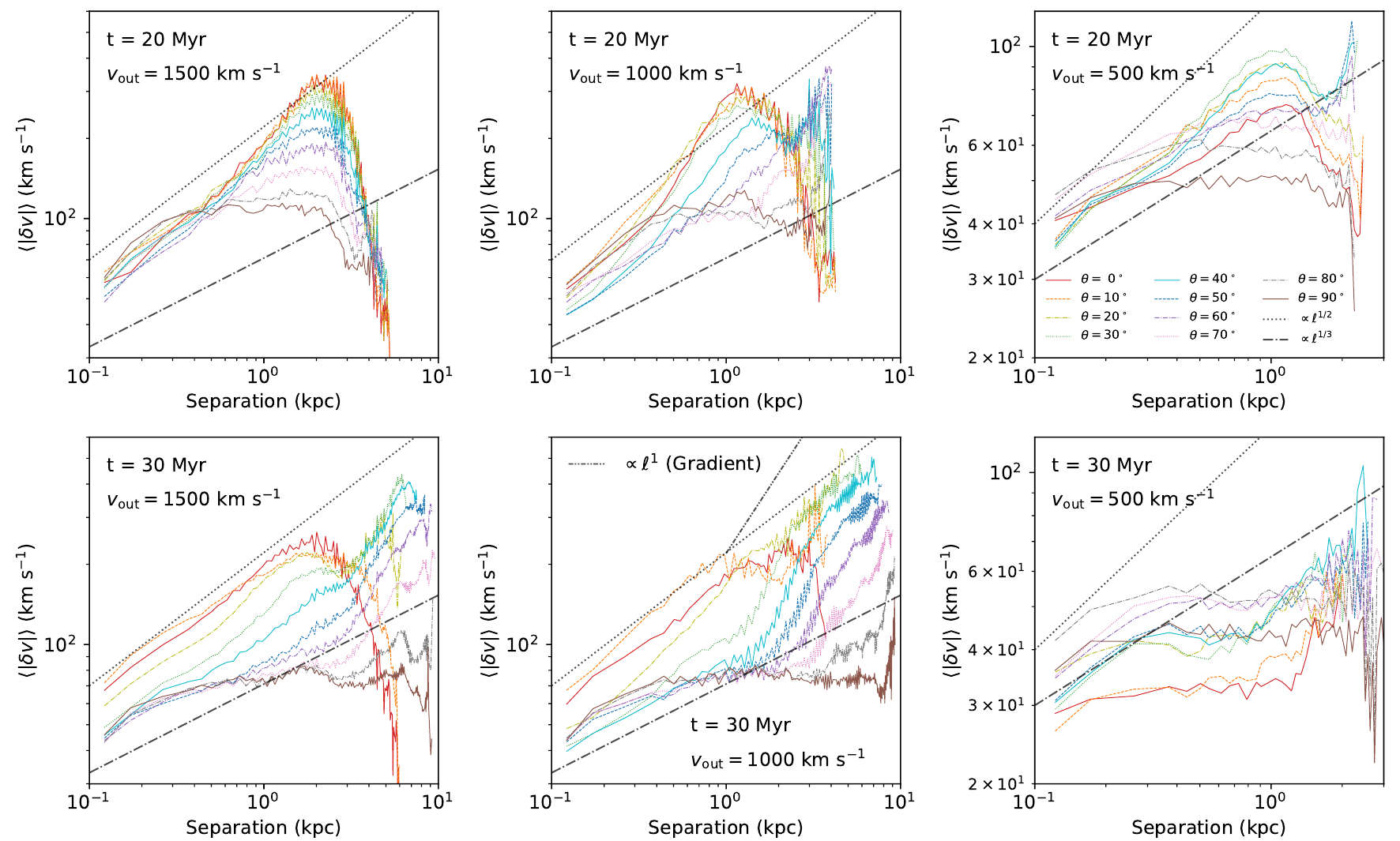}
\caption{Line-of-sight VSF of the simulated cold gas from different viewing angles ($\theta$; between sightline and initial outflow direction, indicated by the different line colors), for outflows with different initial velocities ($v_{\rm out} = 1500,\ 1000,\ 500\,\kms$), at two epochs of evolution ($t=20,\ 30\,$Myr). The scaling relations of $v\propto l^{1/3}$ and $v\propto l^{1/2}$ are plotted for comparison. The lines representing the scaling relations are shifted in each column to approximately bracket the VSF lines. An additional scaling of $v\propto l^1$ is plotted in the lower middle panel, which shows that the VSF is dominated by the velocity gradient at large separations. Note that the cut-off of the lines at separations beyond a few kiloparsecs is constrained by the size of each filament complex, without the contribution from nearby filaments at separations beyond 10\,kpc, as shown in Fig.~\ref{fig:obs}.}
\label{fig:sim}
\end{figure*}

%%%%%%%%%%%%%%%%%%%%%%%%%%%%%%%%%%%%%%%%%%%%%%%%%%
% FIGURE 3 observation
%%%%%%%%%%%%%%%%%%%%%%%%%%%%%%%%%%%%%%%%%%%%%%%%%%
\begin{figure*}[t!]
\centering
\includegraphics[width=\linewidth]{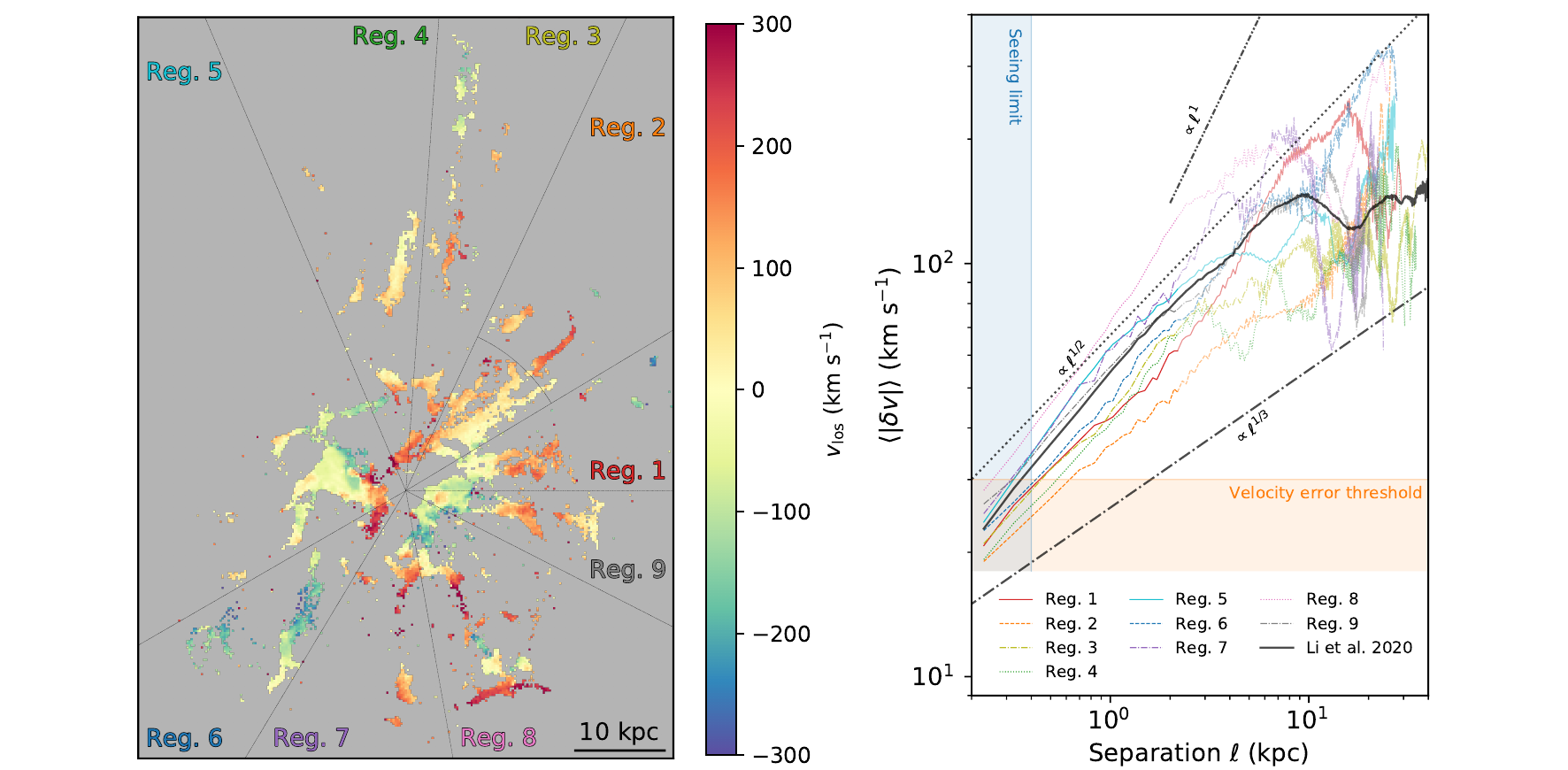}
\caption{Left: Line-of-sight velocity distribution of the H$\alpha$ nebula in the Perseus cluster \citep[][with the systemic velocity shifted to $5264\kms$]{GM2018}, divided into 9 azimuthal regions. Only pixels with velocity errors $<30 \kms$ are plotted and used in the VSF calculation. The arc in Region 2 separates the inner filament complex lacking a radial gradient, and the outer horseshoe filament. An X-ray shock front is detected across Regions 4 and 5, which also indicates supersonic flows \citep{Fabian2003a}. Right: The VSF calculated from pixels in each region. The scaling relations of $v\propto l^{1/3}$ (subsonic), $v\propto l^{1/2}$ (supersonic), and $v\propto l^1$ (gradient) are plotted for comparison. The VSF of the entire nebula from \citet{Li2019d}, which represents an average of all 9 regions, is shown by the solid black line. Data points below the velocity error threshold ($30\kms$) and the seeing limit (0.4\,kpc) are marked with shaded regions.}
\label{fig:obs}
\end{figure*}

Ideally, for supersonic flows where the kinetic energy of the fluid elements may be thermalized, the turbulent velocity structure is expected to follow a steeper slope \citep[e.g., $v\propto l^{1/2}$;][]{Burgers1948,Federrath2013}, due to the partial energy cascade from large to small scales. Therefore, in this work we test this hypothesis by examining the velocity structure of the cold gas in the simulated outflow. In Section~\ref{sec:sim} we describe the simulation parameters and the turbulent structure of the simulated cold gas. In Section~\ref{sec:obs} we examine the turbulent structure of the H$\alpha$ filaments in the Perseus cluster using observations taken with SITELLE \citep{GM2018}. Finally, in Section~\ref{sec:discussion}, we discuss the implications of our work and conclude.

%2==============================================================================%

\section{Simulations of Cold Filaments}\label{sec:sim}

\subsection{Simulation setup}\label{sec:setup}

In order to study the turbulent velocity structure of the cold gas, we perform a set of simulations where cold gas cools and fragments out of an AGN-driven outflow. The simulation setup is similar to that presented in \citet{Qiu2021a}. In each simulation, a spherical gas clump is launched from the center of the cluster, with temperature $T_{\rm out}$, outflow velocity $v_{\rm out}$, and mass $M_{\rm out}$. Initially, the clump is in thermal pressure equilibrium with the surrounding ICM. After launch, the gas in the outflow cools radiatively as it travels in the idealized cluster environment modeled on the Perseus cluster \citep{Qiu2018}, with the highest resolution $\sim60$\,pc. The fluid elements comprise electron, hydrogen and helium species coupled to the non-equilibrium chemistry/cooling solver in the code \texttt{Enzo} \citep{Bryan2014}, which allows us to characterize the H$\alpha$ emission from the recombination process \citep[see also][]{Qiu2019}. We caution that the shock smoothing length of the Zeus solver \citep{Stone1992} employed in this work is around 120\,pc, so we bin the pixels $2\times2$ and only focus on the analyses above this scale.\footnote{An implicit assumption in our choice of resolution is that the gas evolution on scales $\lesssim100$\,pc does not affect the turbulent structure above $\sim100$\,pc, even though the gas turbulence may develop into much smaller scales (see also Section~\ref{sec:VSF}).}

Based on the observational constraints on the cold gas mass, we fix the initial mass $M_{\rm out}=10^8\ms$ of the outflowing gas clump. In the previous simulations, we varied $T_{\rm out}$ around $10^7$\,K, and $v_{\rm out}$ from $1200-2000\kms$ to study the dynamical and morphological evolution of the emergent cold gas and compare it with observed filaments. In this new set of simulations, in order to focus on the supersonic/subsonic nature of the originating outflow, we fix $T_{\rm out}=10^7$\,K, and vary the initial velocity from supersonic to subsonic regime, $v_{\rm out}=1500,\,1000,\ {\rm or}\ 500\kms$. Note that the sound speed for $10^7$\,K plasma is $\approx 500\kms$, so the Mach number of the outflows explored in this work is between 1 and 3. 

For the two cases where $v_{\rm out}=1500,\,1000\kms$, the outflows are launched 1\,kpc away from the cluster center. Due to the slow velocity that inhibits the spatial reach of the outflow, for the last case where $v_{\rm out}=500\kms$ we launch the outflow 10\,kpc away from the cluster center. With this scenario we intend to explore a slower, subsonic phase of the outflow, in a hypothetical scenario where it emerges from the central 10\,kpc with a temperature of $10^7$\,K. Even though this outflow would in reality also be launched from the cluster center as an initially faster and hotter outflow, in this case we focus only on its decelerated, subsonic component, for simplicity (because the properties of a mixture of supersonic and subsonic turbulence may be difficult to interpret). Additionally, this case allows us to test the scenario where filaments originate outside of the cluster core due to thermal instabilities of the hot ICM moving subsonically \citep[e.g.,][]{Li2015,Wang2019}.

\subsection{Cold gas distribution}\label{sec:dist}

As demonstrated in \citet{Qiu2021a}, the radiative cooling time of the $10^7$\,K outflow in initial pressure equilibrium with the ambient ICM is $t_{\rm cool} \lesssim 10$\,Myr. This is a few times larger than the sound crossing time $t_{\rm cross} \equiv \Delta l/c_{\rm s}\approx 2$\, Myr\,($\Delta l$/kpc), where $\Delta l \sim$\,kpc is the typical width of each observed filament complex, and $c_{\rm s}$ is the sound speed. Before $10^4$\,K cold gas forms, a turning point therefore must exist around a few$\times 10^6$\,K, where $t_{\rm cool}<t_{\rm cross}$. In this transition phase, the plasma cools faster than the timescale required for the turbulent eddies to propagate though the length scale $\Delta l$, allowing the emergent cold gas to preserve the ``frozen-in'' turbulent structure. 

In Fig.~\ref{fig:projections}, we show the spatial distribution of the projected H$\alpha$ emissivity, as well as the emissivity-weighted line-of-sight velocity for the three velocity cases at three evolutionary epochs. In all cases, the cold gas fragments continuously out of the radiatively cooling outflow after $\sim10$\,Myr, forming a filamentary trail as the outflow rises in the cluster potential \citep{Qiu2019,Qiu2020}. As discussed in \citet{Qiu2021a}, the morphology of the cold gas may take both longitudinal and transverse shapes depending on the initial outflow properties, such as the outer horseshoe-shaped shell at $t=20$\,Myr for $v_{\rm out}=1500\kms$. Note that in the slowest outflow case, it takes much longer for the low-temperature plasma to stretch and form elongated cold gas filaments under the effect of cluster gravity, unless the initial thermally unstable region is already spatially extended. We refer the readers to our previous work on the dynamical and morphological evolution of the cold gas, which can be described reasonably well by a 1D model comprising radiative cooling and ICM ram pressure \citep{Qiu2020, Qiu2021a}. In this work, we instead focus on the small-scale turbulent structure that can be extracted from the cold gas velocity.

\subsection{Turbulent velocity structure}\label{sec:VSF}

In order to characterize the turbulent structure of the cold gas in the simulations, in Fig.~\ref{fig:sim} we compute the first-order velocity structure function (VSF). The simulated outflow is first projected along different viewing angles ($\theta$) with respect to the initial outflow direction to obtain both the line-of-sight velocity ($v_{\rm los}$) map and the H$\alpha$ emissivity map (the case where $\theta=90^\circ$ is shown in Fig.~\ref{fig:projections}). $v_{\rm los}$ is weighted by the H$\alpha$ emissivity in each line integral. After projection, pixels from the $v_{\rm los}$ map are selected to compute the VSF if the corresponding projected H$\alpha$ emissivity is $>4\times10^{-5}\ergs\,{\rm cm}^{-2}$, a threshold consistent with observations of the Perseus cluster filaments \citep[][assuming isotropic emission from the redshift $z=0.01756$ of the Perseus cluster]{GM2018}. For all possible pair combinations of the selected pixels, we compute the $v_{\rm los}$ difference $\delta v$ and the projected separation $l$. We then divide the pixel pairs into $\Delta l \approx 0.1$\,kpc bins based on their separation $l$, before taking the average of the absolute velocity difference, $\langle |\delta v^n| \rangle=\langle | v_i-v_j|^n \rangle$ ($n=1$ for the first-order VSF) in each separation bin. Bins with less than 10 pixel pairs are removed from the final plot.

In Fig.~\ref{fig:sim}, we present the VSF of the simulated cold gas with different initial velocities ($v_{\rm out}=1500,\,1000,\ {\rm or}\ 500\kms$), for projections along varying viewing angles $\theta$, at two epochs of evolution ($t=20,\,30$\,Myr). The timestamps represent $\Delta t\approx10,\, 20 $\,Myr after the cold gas formation. The driving scale of the turbulence, where the VSF peaks or flattens, is on the order of a few kpc, corresponding to the size of the initial outflow with coherent velocity (which later develops into varying speeds on smaller scales due to turbulence). In all cases at the early epoch ($t=20$\,Myr), the VSF below 1\,kpc follows the power indices ($\gamma$) expected of super- and sub-sonic turbulence at small-to-intermediate viewing angles ($\theta\lesssim80^\circ$), i.e., $\gamma\approx1/2$ for $v_{\rm out}=1500,\,1000\,\kms$, and $\gamma\approx1/3$ for $v_{\rm out}=500\kms$. At these viewing angles, the line-of-sight velocity ($v_{\rm los}$) is dominated by the component parallel to $v_{\rm out}$, which is the driving direction of turbulence. For small $\theta\lesssim 30^\circ$, the slope slightly steepens for all cases around 1\,kpc, when the viewing angle is close to the outflow direction. This may be contributed by the steep velocity gradient for small $\theta$, which features a scaling relation $v\propto l$, as discussed in more detail in Section~\ref{sec:obs} below. On the other hand, for large $\theta\gtrsim80^\circ$, $\gamma$ flattens to 0 because the $v_{\rm los}$ is nearly perpendicular to the outflow direction, which results in a $v_{\rm los}$ distribution similar at all separation scales (e.g., the $v_{\rm los}$ distribution shown in Fig.~\ref{fig:projections}). For the average viewing angle, however, the power index $\gamma$ of the cold gas VSF below a few kiloparsecs signifies the property of the plasma from which the filaments originate, i.e., supersonic for $\gamma\approx 1/2$, or subsonic for $\gamma\approx1/3$.

After the cold clumps fragment out of the plasma, their interaction is primarily gravitational, both with the background potential and with each other. This inviscid interaction inevitably hinders the energy cascade and leads to the flattening of the velocity structure at smaller scales. In the simulations, this trend gradually develops after $t=20$\,Myr (or 10\,Myr after cold gas formation). Note that the sound speed for the cold clumps below $10^4$\,K is $c_{\rm s}\lesssim 10\kms$, indicating that the gas may still be traveling supersonically. However, the turbulent structure likely develops at scales much smaller than those probed both in our simulations and in observations of galaxy clusters \citep[for an example of the turbulence on parsec scales, see the velocity structure and the small-scale flattening probed by the stellar velocity in the Orion Complex;][]{Ha2021}.

%%%%%%%%%%%%%%%%%%%%%%%%%%%%%%%%%%%%%%%%%%%%%%%%%%
% FIGURE 4 gradient
%%%%%%%%%%%%%%%%%%%%%%%%%%%%%%%%%%%%%%%%%%%%%%%%%%
\begin{figure*}[t!]
\centering
\includegraphics[width=\linewidth]{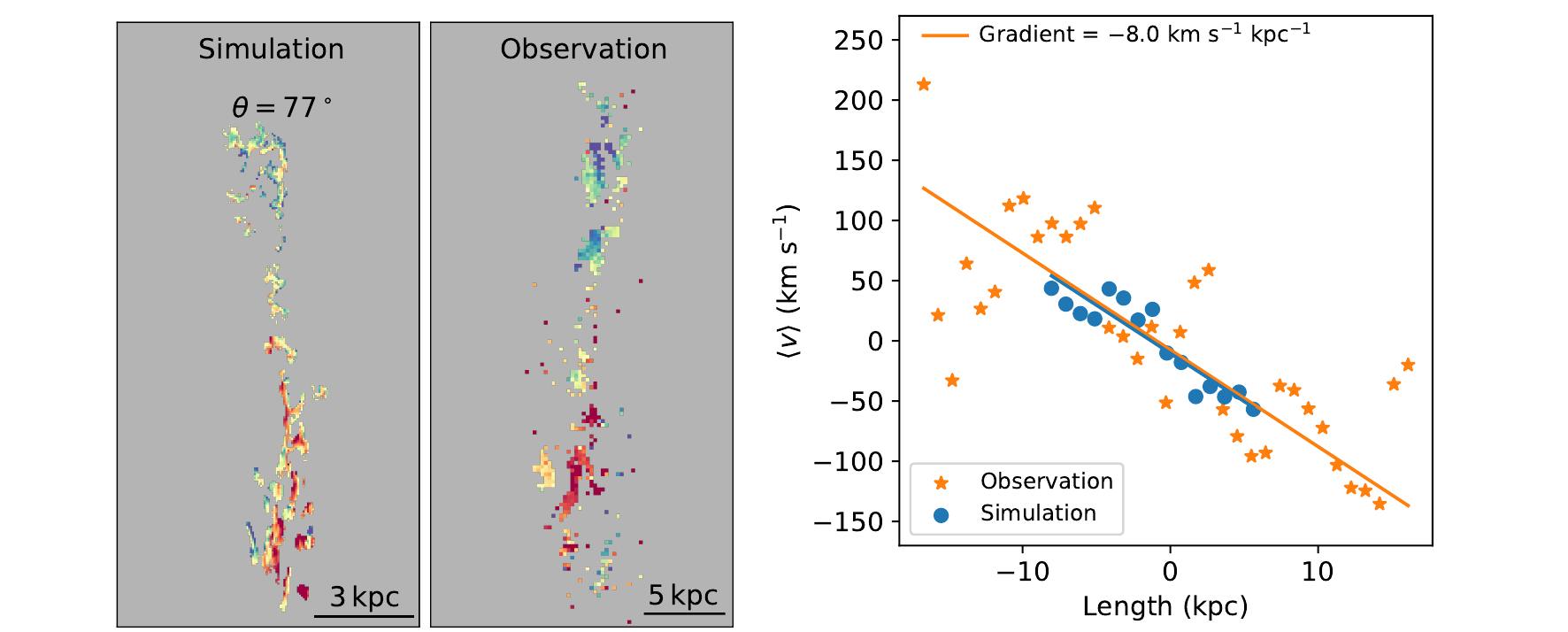}
\caption{Left two panels: Line-of-sight velocity distribution of the simulated filament ($v_{\rm out}=1000\kms$, $t=40$\,Myr, $\theta=77^\circ$) and the observed northern filament in the Perseus cluster. Right: Comparison of the velocity gradient over the length of the simulated and observed filaments ($-8.0\kms\,{\rm kpc}^{-1}$). Data points show the flux-weighted average velocity in each length bin of 1\,kpc. Solid lines show the linear fit to the data sets. Points with deviations larger than $0.5\,\sigma$ from the fit are omitted, which removes bins with few pixels at both ends and the middle of the observed filament.}
\label{fig:gradient}
\end{figure*}

This trend of flattened VSF is seen in the later epoch of the simulated cold gas. In the lower panels of Fig.~\ref{fig:sim}, where some cold gas clumps have existed for more than 20\,Myr, their relative motion has ceased developing into smaller scales, e.g., forming binary-like systems. The VSF flattens significantly compared with the early epoch, with $\gamma$ ranging between 1/3 and 1/2 in the cases where $v_{\rm out}=1500,\,1000\kms$ ($\gamma$ decreases with larger $\theta$), and with $\gamma$ approaching 0 in the $v_{\rm out}=500\kms$ case, at separations between 0.3 and 1\,kpc. Realistically, however, the cold gas will likely continue to form molecules or fuel star formation, dropping out of the gas phase probed by H$\alpha$ emission. Therefore, the lack of $\gamma\leq 1/3$ in observations (see Fig.~\ref{fig:obs}) indicates that the lifetime of the H$\alpha$ emitting gas should be shorter than 20\,Myr, before gravitational interactions flatten the VSF. 

%3==============================================================================%
\section{Comparison between Simulated and Observed Filaments in Perseus}\label{sec:obs}

In this section we examine filament observations in the Perseus cluster taken with the optical imaging Fourier transform spectrometer SITELLE at the Canada-France-Hawaii Telescope \citep[CFHT;][]{GM2018}. These data were taken at a spectral resolution of $R=1800$ in a filter covering $647-685$\,nm. SITELLE's angular resolution is 0.321''$\times$0.321'', and the data cubes are binned by a factor of $2\times2$ (to an effective pixel resolution of 0.642''$\times$0.642'') to increase the signal-to-noise ratio. Three lines ([NII]$\lambda$6548, H$\alpha$, and [NII]$\lambda$6584) were simultaneously fit in each pixel using Gaussian functions (convolved with the instrument line shape), with typical central velocity shift errors between $5\kms$ and $20\kms$. Compared with the map published in \citet{GM2018}, this map includes several small faint filaments that were previously cut off in the west (right) region. The same H$\alpha$ flux cut ($3\times10^{-17}\,{\rm erg\,s}^{-1}\,{\rm cm}^{-2}\,{\rm pixel}^{-1}$) is applied to the maps for the analysis.

The steeper slope of the cold gas VSF in the Perseus cluster was first discovered by \citet{Li2019d}, who examined the entire filament nebula and found a peak in the VSF at $\sim10$\,kpc, indicating that the filaments may be lifted by the AGN-inflated X-ray cavities \citep[e.g.,][]{Revaz2008, McNamara2016}. However, unlike AGN jets or winds that drive supersonic flows \citep{Qiu2020, Hillel2019}, the buoyantly rising cavities often travel at subsonic speeds in the centers of galaxy clusters \citep[see, e.g., the bubble speed estimates in][]{Birzan2004}, unlikely to display steeper slopes without additional damping mechanisms such as magnetic fields \citep[e.g.,][]{Wang2020, Mohapatra2021a}. In order to test the AGN-driven scenario, we further examine the turbulent structure of the observed filaments and compare it with the outflow model in our simulations. Noting that the entire nebula consists of multiple generations of filaments driven by past AGN activity at different epochs, each filament complex should retain structural information about its origin. Therefore, we examine the VSF of filament complexes divided into 9 azimuthal regions in Fig.~\ref{fig:obs} to compare with our supersonic outflow model. We note that precisely separating individual outflows is often challenging due to projection effects and overlapping filaments. Therefore, this division represents a simple approach to probe the turbulence likely driven by the same AGN outburst in a particular direction. We also caution that the seeing limit of the observation is about 1.1'' ($\approx0.4$\,kpc, FWHM), so we focus our comparison above this length scale.

In Fig.~\ref{fig:obs}, we plot the VSFs for the azimuthal regions, each showing a steep slope with $\gamma\approx1/2$ at scales $\lesssim 2$kpc. The steep slopes indicate that the turbulence in these filaments is supersonic, in agreement with the supersonic outflows modeled in our simulations. The only exception occurs in Region 2, where $1/3< \gamma <1/2$.\footnote{Note there is a flattened ``ghost'' X-ray cavity located near the outer horseshoe filament, which may suggest that the filament is lifted by the rising bubble \citep{Fabian2003}. We have further separated the inner and outer filaments in Region 2, as indicated by the arc in Figure~\ref{fig:obs}, and found that the VSF slope of the horseshoe filament is still $1/2$. The flattening of the VSF in Region 2 is therefore due to the inner filament complex.} The inner filament complex (below the arc located at $r=19$\,kpc in Fig.~\ref{fig:obs}) in this region lacks a radial gradient, indicating that the viewing angle $\theta$ may be perpendicular to the direction of motion, therefore the VSF may be contaminated by the uniform $v_{\rm los}$ distribution discussed in Section~\ref{sec:sim}. Nevertheless, the VSF slopes below the driving scale of $\sim2$\,kpc indicate the ubiquity of supersonic turbulence in the Perseus cluster. Compared with the simulation results presented in this work, we find that the supersonic turbulence structure is a direct indicator that the cold gas fragmented out of a fast outflow driven by recent AGN activity. 

Beyond the driving scale of $\sim 2$\,kpc, however, the VSF depends primarily on the radial gradient of the filament complex in each region, which has a characteristic slope of 1, as indicated by the linear relation $v\propto l$ in the right panel of Fig.~\ref{fig:obs}. For Regions 1 and 6 with large radial gradients, the VSF continues to rise, while for Region 4 with a small radial gradient, the VSF drops immediately. The average behavior, however, leads to the extension of the $\gamma\approx1/2$ slope from $2-10$\,kpc noted in \citet{Li2019d}. As a case study, the velocity gradient of the northern filament in Region 3 is calculated in Fig.~\ref{fig:gradient}. While the gradient of $-8.0\kms\,{\rm kpc}^{-1}$ does not significantly contribute to the VSF below $\sim2$\,kpc, it yields a velocity difference of $80\kms$ at the separation of 10\,kpc, comparable to the VSF amplitude in Region 3 ($\sim100\kms$ at 10\,kpc). In the case of Region 8, the mixed radial gradient, likely due to overlapping filaments at different epochs of evolution, steepens the VSF slope significantly at separations near $\sim1$\,kpc. We therefore caution that the turbulent VSF slope shall only be extracted at spatial separations minimally affected by the radial velocity gradient.
%3.5 ==============================================================================%

Given the dependence of $v_{\rm los}$ on the viewing angle $\theta$, we also try to constrain $\theta$ for the most extended northern filament in Region 3 of the Perseus cluster. As shown in Fig.~\ref{fig:gradient}, there is a smooth velocity gradient along the length of the northern filament, with the bottom half redshifted and the top half blueshifted. A similar morphology and velocity gradient can be found in the simulated filament with $v_{\rm out} = 1000\kms$ at $t=40$\,Myr. Compared with the simulation, the mixed velocity shifts indicate that the bottom half contains an older generation of gas that fragmented out of the outflow and has started to fall back, while the gas in the top half is still rising in the cluster potential. The same line-of-sight velocity gradient can be reproduced in the simulation when the viewing angle $\theta=77^\circ$. If the northern filament is similarly inclined, the angle indicates a factor of $1/\sin\theta\approx1.03$ increase to the filament extent, and a factor of $1/\cos \theta\approx4.4$ increase to the velocity. This boosts the maximum inflow/outflow speeds to $\approx900\kms$, significantly larger than those commonly observed of the filaments.

%4 ==============================================================================%

\section{Conclusions}\label{sec:discussion}

%summary
In this paper we perform hydrodynamical simulations to study the turbulent velocity structure of the cold gas that fragments out of radiatively cooling hot outflows. By varying the initial velocities, we find that the cold gas turbulent velocity structure depends on the supersonic or subsonic nature of the original outflow, and can be used in observations to probe the interactions between AGN and the ICM. The main findings are summarized below: 

%results
1. For hot outflows with a short radiative cooling timescale $t_{\rm cool}$, the turbulent velocity structure may be ``frozen'' in the emerging cold gas, when the dynamical sound crossing time $t_{\rm cross}>t_{\rm cool}$. The preserved velocity structure therefore can be an indicator of the characteristic velocity of the plasma out of which the cold gas fragments. For viewing angles $\theta\lesssim80^\circ$ with respect to the outflow direction, the slope of the first-order velocity structure function is $\geq1/2$ for the cold gas that originates from supersonic flows, or $\approx1/3$ for subsonic flows. 

2. For individual filaments in both the simulations and the observations of the Perseus cluster, the VSF slope extends from a few hundred parsecs up to the driving scale of a few kiloparsecs, corresponding to the size of the original outflow, as well as the width of the filament complex. Both the VSF slope and the driving scale can therefore be used to constrain properties of the AGN-driven outflows.

3. Beyond the driving scale, the VSF slope is contributed primarily by the velocity gradient along the length of the filament. Because the projected gradient depends on the viewing angle $\theta$, this contribution may steepen the VSF slope for small $\theta$ (large gradient) at separations near $\sim 1$\,kpc. By varying $\theta$ of the simulated filament to compare with the velocity gradient of the most extended northern filament in the Perseus cluster, we infer that the viewing angle of the northern filament is $\approx77^\circ$. This angle implies that the maximum inflow/outflow speed in the northern filament is $\approx900\kms$, which corroborates the argument that supersonic flows are needed to reproduce the observed turbulent structure of the filaments.

4. Gravitational interactions of the cold gas result in the flattening of the VSF slope in the simulations over 20\,Myr, which indicates the stagnation of the energy cascade without additional mechanisms to transfer the kinetic energy or dissipate the cold gas. The lack of flattened VSF in the observed Perseus filaments suggests that the H$\alpha$-emitting gas is short-lived with a lifespan $<20$\,Myr, during which time it either fuels the molecular gas growth and the scattered, intracluster star formation, or becomes recycled to the ICM by local heating processes.

We note that while the filament VSF provides a powerful diagnostic tool for understanding the origin of the filaments and the dynamic interaction between the AGN and the ICM, (a) the supersonic/subsonic nature of the originating outflow, (b) the steepening from velocity gradient, (c) the flattening due to gravitational interactions, as well as (d) the modulation by the viewing angle all contribute to the final VSF of the filaments. The modeling presented in this work therefore provides a viable method to disentangle these (sometimes competing) parameters for observed filaments. Through the modeling of the dynamical evolution of different generations of filaments, the simulations can also constrain the duty cycle and reconstruct the AGN activity in the recent few$\times10$\,Myr, at the cores of galaxy clusters.

%%%%%%%%%%%%%%%%%%%%%%%%%%%%%%%%%%%%%%%%%%%%%%%%%%%%%%%%%%%%%%%%%%%%%%%%%%%%%%%%%%%%%%%%%%%%%%%%%%%%%%%%%%%

%\hspace{1cm}
\begin{acknowledgements}
\noindent This work is supported by the National Natural Science Foundation of China (12003003, 12073003, 11721303, 11991052, 11950410493), the China Postdoctoral Science Foundation (2020T130019), the National Key R\&D Program of China (2016YFA0400702), and the High-Performance Computing Platform of Peking University. M.G. acknowledges financial support from grant RTI2018-096228-B-C31 (MCIU/AEI/FEDER, UE), from the coordination of the participation in SKA-SPAIN financed by the Ministry of Science and Innovation (MICIN), and from the State Agency for Research of the Spanish Ministry of Science, Innovation and Universities through the ``Center of Excellence Severo Ochoa'' awarded to the Instituto de Astrof\'isica de Andaluc\'ia (SEV-2017-0709).
\end{acknowledgements}

%%%%%%%%%%%%%%%%%%%%%%%%%%%%%%%%%%%%%%%%%%%%%%%%%%%%%%%%%%%%%%%%%%%%%%%%%%%%%%%%%%%%%%%%%%%%%%%%%%%%%%%%%%%

%\bibliography{bibtex}{}

\begin{thebibliography}{}
\expandafter\ifx\csname natexlab\endcsname\relax\def\natexlab#1{#1}\fi
\providecommand{\url}[1]{\href{#1}{#1}}
\providecommand{\dodoi}[1]{doi:~\href{http://doi.org/#1}{\nolinkurl{#1}}}
\providecommand{\doeprint}[1]{\href{http://ascl.net/#1}{\nolinkurl{http://ascl.net/#1}}}
\providecommand{\doarXiv}[1]{\href{https://arxiv.org/abs/#1}{\nolinkurl{https://arxiv.org/abs/#1}}}

\bibitem[{{B{\^\i}rzan} {et~al.}(2004){B{\^\i}rzan}, {Rafferty}, {McNamara},
  {Wise}, \& {Nulsen}}]{Birzan2004}
{B{\^\i}rzan}, L., {Rafferty}, D.~A., {McNamara}, B.~R., {Wise}, M.~W., \&
  {Nulsen}, P.~E.~J. 2004, \apj, 607, 800, \dodoi{10.1086/383519}

\bibitem[{{Bryan} {et~al.}(2014){Bryan}, {Norman}, {O'Shea}, {Abel}, {Wise},
  {Turk}, {Reynolds}, {Collins}, {Wang}, {Skillman}, {Smith}, {Harkness},
  {Bordner}, {Kim}, {Kuhlen}, {Xu}, {Goldbaum}, {Hummels}, {Kritsuk}, {Tasker},
  {Skory}, {Simpson}, {Hahn}, {Oishi}, {So}, {Zhao}, {Cen}, {Li}, \& {Enzo
  Collaboration}}]{Bryan2014}
{Bryan}, G.~L., {Norman}, M.~L., {O'Shea}, B.~W., {et~al.} 2014, \apjs, 211,
  19, \dodoi{10.1088/0067-0049/211/2/19}

\bibitem[{{Burgers}(1948)}]{Burgers1948}
{Burgers}, J. 1948, Advances in Applied Mechanics, 171,
  \dodoi{10.1016/S0065-2156(08)70100-5}

\bibitem[{{Conselice} {et~al.}(2001){Conselice}, {Gallagher}, \&
  {Wyse}}]{Conselice2001}
{Conselice}, C.~J., {Gallagher}, John~S., I., \& {Wyse}, R. F.~G. 2001, \aj,
  122, 2281, \dodoi{10.1086/323534}

\bibitem[{{Fabian}(2012)}]{Fabian2012}
{Fabian}, A.~C. 2012, \araa, 50, 455,
  \dodoi{10.1146/annurev-astro-081811-125521}

\bibitem[{{Fabian} {et~al.}(2003{\natexlab{a}}){Fabian}, {Sanders}, {Allen},
  {Crawford}, {Iwasawa}, {Johnstone}, {Schmidt}, \& {Taylor}}]{Fabian2003a}
{Fabian}, A.~C., {Sanders}, J.~S., {Allen}, S.~W., {et~al.} 2003{\natexlab{a}},
  \mnras, 344, L43, \dodoi{10.1046/j.1365-8711.2003.06902.x}

\bibitem[{{Fabian} {et~al.}(2003{\natexlab{b}}){Fabian}, {Sanders}, {Crawford},
  {Conselice}, {Gallagher}, \& {Wyse}}]{Fabian2003}
{Fabian}, A.~C., {Sanders}, J.~S., {Crawford}, C.~S., {et~al.}
  2003{\natexlab{b}}, \mnras, 344, L48,
  \dodoi{10.1046/j.1365-8711.2003.06856.x}

\bibitem[{{Federrath}(2013)}]{Federrath2013}
{Federrath}, C. 2013, \mnras, 436, 1245, \dodoi{10.1093/mnras/stt1644}

\bibitem[{{Gendron-Marsolais} {et~al.}(2018){Gendron-Marsolais},
  {Hlavacek-Larrondo}, {Martin}, {Drissen}, {McDonald}, {Fabian}, {Edge},
  {Hamer}, {McNamara}, \& {Morrison}}]{GM2018}
{Gendron-Marsolais}, M., {Hlavacek-Larrondo}, J., {Martin}, T.~B., {et~al.}
  2018, \mnras, 479, L28, \dodoi{10.1093/mnrasl/sly084}

\bibitem[{{Ha} {et~al.}(2021){Ha}, {Li}, {Xu}, {Kounkel}, \& {Li}}]{Ha2021}
{Ha}, T., {Li}, Y., {Xu}, S., {Kounkel}, M., \& {Li}, H. 2021, \apjl, 907, L40,
  \dodoi{10.3847/2041-8213/abd8c9}

\bibitem[{{Hillel} \& {Soker}(2020)}]{Hillel2019}
{Hillel}, S., \& {Soker}, N. 2020, \apj, 896, 104,
  \dodoi{10.3847/1538-4357/ab9109}

\bibitem[{{Hitomi Collaboration} {et~al.}(2016){Hitomi Collaboration},
  {Aharonian}, {Akamatsu}, {Akimoto}, {Allen}, {Anabuki}, {Angelini}, {Arnaud},
  {Audard}, {Awaki}, {Axelsson}, {Bamba}, {Bautz}, {Blandford}, {Brenneman},
  {Brown}, {Bulbul}, {Cackett}, {Chernyakova}, {Chiao}, {Coppi}, {Costantini},
  {de Plaa}, {den Herder}, {Done}, {Dotani}, {Ebisawa}, {Eckart}, {Enoto},
  {Ezoe}, {Fabian}, {Ferrigno}, {Foster}, {Fujimoto}, {Fukazawa}, {Furuzawa},
  {Galeazzi}, {Gallo}, {Gandhi}, {Giustini}, {Goldwurm}, {Gu}, {Guainazzi},
  {Haba}, {Hagino}, {Hamaguchi}, {Harrus}, {Hatsukade}, {Hayashi}, {Hayashi},
  {Hayashida}, {Hiraga}, {Hornschemeier}, {Hoshino}, {Hughes}, {Iizuka},
  {Inoue}, {Inoue}, {Ishibashi}, {Ishida}, {Ishikawa}, {Ishisaki}, {Itoh},
  {Iyomoto}, {Kaastra}, {Kallman}, {Kamae}, {Kara}, {Kataoka}, {Katsuda},
  {Katsuta}, {Kawaharada}, {Kawai}, {Kelley}, {Khangulyan}, {Kilbourne},
  {King}, {Kitaguchi}, {Kitamoto}, {Kitayama}, {Kohmura}, {Kokubun}, {Koyama},
  {Koyama}, {Kretschmar}, {Krimm}, {Kubota}, {Kunieda}, {Laurent}, {Lebrun},
  {Lee}, {Leutenegger}, {Limousin}, {Loewenstein}, {Long}, {Lumb}, {Madejski},
  {Maeda}, {Maier}, {Makishima}, {Markevitch}, {Matsumoto}, {Matsushita},
  {McCammon}, {McNamara}, {Mehdipour}, {Miller}, {Miller}, {Mineshige},
  {Mitsuda}, {Mitsuishi}, {Miyazawa}, {Mizuno}, {Mori}, {Mori}, {Moseley},
  {Mukai}, {Murakami}, {Murakami}, {Mushotzky}, {Nagino}, {Nakagawa},
  {Nakajima}, {Nakamori}, {Nakano}, {Nakashima}, {Nakazawa}, {Nobukawa},
  {Noda}, {Nomachi}, {O'Dell}, {Odaka}, {Ohashi}, {Ohno}, {Okajima}, {Ota},
  {Ozaki}, {Paerels}, {Paltani}, {Parmar}, {Petre}, {Pinto}, {Pohl}, {Porter},
  {Pottschmidt}, {Ramsey}, {Reynolds}, {Russell}, {Safi-Harb}, {Saito},
  {Sakai}, {Sameshima}, {Sato}, {Sato}, {Sato}, {Sawada}, {Schartel},
  {Serlemitsos}, {Seta}, {Shidatsu}, {Simionescu}, {Smith}, {Soong}, {Stawarz},
  {Sugawara}, {Sugita}, {Szymkowiak}, {Tajima}, {Takahashi}, {Takahashi},
  {Takeda}, {Takei}, {Tamagawa}, {Tamura}, {Tamura}, {Tanaka}, {Tanaka},
  {Tanaka}, {Tashiro}, {Tawara}, {Terada}, {Terashima}, {Tombesi}, {Tomida},
  {Tsuboi}, {Tsujimoto}, {Tsunemi}, {Tsuru}, {Uchida}, {Uchiyama}, {Uchiyama},
  {Ueda}, {Ueda}, {Ueno}, {Uno}, {Urry}, {Ursino}, {de Vries}, {Watanabe},
  {Werner}, {Wik}, {Wilkins}, {Williams}, {Yamada}, {Yamaguchi}, {Yamaoka},
  {Yamasaki}, {Yamauchi}, {Yamauchi}, {Yaqoob}, {Yatsu}, {Yonetoku}, {Yoshida},
  {Yuasa}, {Zhuravleva}, \& {Zoghbi}}]{Hitomi2016}
{Hitomi Collaboration}, {Aharonian}, F., {Akamatsu}, H., {et~al.} 2016, \nat,
  535, 117, \dodoi{10.1038/nature18627}

\bibitem[{{Kolmogorov}(1941)}]{Kolmogorov1941}
{Kolmogorov}, A.~N. 1941, Akademiia Nauk SSSR Doklady, 32, 16

\bibitem[{{Li} {et~al.}(2015){Li}, {Bryan}, {Ruszkowski}, {Voit}, {O'Shea}, \&
  {Donahue}}]{Li2015}
{Li}, Y., {Bryan}, G.~L., {Ruszkowski}, M., {et~al.} 2015, \apj, 811, 73,
  \dodoi{10.1088/0004-637X/811/2/73}

\bibitem[{{Li} {et~al.}(2020){Li}, {Gendron-Marsolais}, {Zhuravleva}, {Xu},
  {Simionescu}, {Tremblay}, {Lochhaas}, {Bryan}, {Quataert}, {Murray},
  {Boselli}, {Hlavacek-Larrondo}, {Zheng}, {Fossati}, {Li}, {Emsellem},
  {Sarzi}, {Arzamasskiy}, \& {Vishniac}}]{Li2019d}
{Li}, Y., {Gendron-Marsolais}, M.-L., {Zhuravleva}, I., {et~al.} 2020, \apjl,
  889, L1, \dodoi{10.3847/2041-8213/ab65c7}

\bibitem[{{McNamara} \& {Nulsen}(2012)}]{McNamara2012}
{McNamara}, B.~R., \& {Nulsen}, P.~E.~J. 2012, New Journal of Physics, 14,
  055023, \dodoi{10.1088/1367-2630/14/5/055023}

\bibitem[{{McNamara} {et~al.}(2016){McNamara}, {Russell}, {Nulsen}, {Hogan},
  {Fabian}, {Pulido}, \& {Edge}}]{McNamara2016}
{McNamara}, B.~R., {Russell}, H.~R., {Nulsen}, P.~E.~J., {et~al.} 2016, \apj,
  830, 79, \dodoi{10.3847/0004-637X/830/2/79}

\bibitem[{{Mohapatra} {et~al.}(2022){Mohapatra}, {Jetti}, {Sharma}, \&
  {Federrath}}]{Mohapatra2021a}
{Mohapatra}, R., {Jetti}, M., {Sharma}, P., \& {Federrath}, C. 2022, \mnras,
  510, 2327, \dodoi{10.1093/mnras/stab3429}

\bibitem[{{Qiu} {et~al.}(2019{\natexlab{a}}){Qiu}, {Bogdanovi{\'c}}, {Li}, \&
  {McDonald}}]{Qiu2019}
{Qiu}, Y., {Bogdanovi{\'c}}, T., {Li}, Y., \& {McDonald}, M.
  2019{\natexlab{a}}, \apjl, 872, L11, \dodoi{10.3847/2041-8213/ab0375}

\bibitem[{{Qiu} {et~al.}(2020){Qiu}, {Bogdanovi{\'c}}, {Li}, {McDonald}, \&
  {McNamara}}]{Qiu2020}
{Qiu}, Y., {Bogdanovi{\'c}}, T., {Li}, Y., {McDonald}, M., \& {McNamara}, B.~R.
  2020, Nature Astronomy, 4, 900, \dodoi{10.1038/s41550-020-1090-7}

\bibitem[{{Qiu} {et~al.}(2019{\natexlab{b}}){Qiu}, {Bogdanovi{\'c}}, {Li},
  {Park}, \& {Wise}}]{Qiu2018}
{Qiu}, Y., {Bogdanovi{\'c}}, T., {Li}, Y., {Park}, K., \& {Wise}, J.~H.
  2019{\natexlab{b}}, \apj, 877, 47, \dodoi{10.3847/1538-4357/ab18fd}

\bibitem[{{Qiu} {et~al.}(2021{\natexlab{a}}){Qiu}, {Hu}, {Inayoshi}, {Ho},
  {Bogdanovi{\'c}}, \& {McNamara}}]{Qiu2021a}
{Qiu}, Y., {Hu}, H., {Inayoshi}, K., {et~al.} 2021{\natexlab{a}}, \apjl, 917,
  L7, \dodoi{10.3847/2041-8213/ac16d9}

\bibitem[{{Qiu} {et~al.}(2021{\natexlab{b}}){Qiu}, {McNamara},
  {Bogdanovi{\'c}}, {Inayoshi}, \& {Ho}}]{Qiu2021}
{Qiu}, Y., {McNamara}, B.~R., {Bogdanovi{\'c}}, T., {Inayoshi}, K., \& {Ho},
  L.~C. 2021{\natexlab{b}}, \apj, 923, 256, \dodoi{10.3847/1538-4357/ac2ede}

\bibitem[{{Revaz} {et~al.}(2008){Revaz}, {Combes}, \& {Salom{\'e}}}]{Revaz2008}
{Revaz}, Y., {Combes}, F., \& {Salom{\'e}}, P. 2008, \aap, 477, L33,
  \dodoi{10.1051/0004-6361:20078915}

\bibitem[{{Sanders} {et~al.}(2011){Sanders}, {Fabian}, \&
  {Smith}}]{Sanders2011}
{Sanders}, J.~S., {Fabian}, A.~C., \& {Smith}, R.~K. 2011, \mnras, 410, 1797,
  \dodoi{10.1111/j.1365-2966.2010.17561.x}

\bibitem[{{Shi} {et~al.}(2021){Shi}, {Li}, {Yuan}, \& {Zhu}}]{Shi2021}
{Shi}, F., {Li}, Z., {Yuan}, F., \& {Zhu}, B. 2021, Nature Astronomy, 5, 928,
  \dodoi{10.1038/s41550-021-01394-0}

\bibitem[{{Stone} {et~al.}(1992){Stone}, {Mihalas}, \& {Norman}}]{Stone1992}
{Stone}, J.~M., {Mihalas}, D., \& {Norman}, M.~L. 1992, \apjs, 80, 819,
  \dodoi{10.1086/191682}

\bibitem[{{Wang} {et~al.}(2021){Wang}, {Ruszkowski}, {Pfrommer}, {Oh}, \&
  {Yang}}]{Wang2020}
{Wang}, C., {Ruszkowski}, M., {Pfrommer}, C., {Oh}, S.~P., \& {Yang}, H. Y.~K.
  2021, \mnras, 504, 898, \dodoi{10.1093/mnras/stab966}

\bibitem[{{Wang} {et~al.}(2020){Wang}, {Ruszkowski}, \& {Yang}}]{Wang2019}
{Wang}, C., {Ruszkowski}, M., \& {Yang}, H. Y.~K. 2020, \mnras, 493, 4065,
  \dodoi{10.1093/mnras/staa550}

\bibitem[{{Zhuravleva} {et~al.}(2019){Zhuravleva}, {Churazov}, {Schekochihin},
  {Allen}, {Vikhlinin}, \& {Werner}}]{Zhuravleva2019}
{Zhuravleva}, I., {Churazov}, E., {Schekochihin}, A.~A., {et~al.} 2019, Nature
  Astronomy, 3, 832, \dodoi{10.1038/s41550-019-0794-z}

\bibitem[{{Zhuravleva} {et~al.}(2014){Zhuravleva}, {Churazov}, {Schekochihin},
  {Allen}, {Ar{\'e}valo}, {Fabian}, {Forman}, {Sanders}, {Simionescu},
  {Sunyaev}, {Vikhlinin}, \& {Werner}}]{Zhuravleva2014}
---. 2014, \nat, 515, 85, \dodoi{10.1038/nature13830}

\end{thebibliography}
%\bibliographystyle{aasjournal}

\end{CJK*}

\end{document}